# A Mixed-Methods Analysis of the Algorithm-Mediated Labor of Online Food Deliverers in China

ZHILONG CHEN, Department of Electronic Engineering, Tsinghua University, China
XIAOCHONG LAN, Department of Electronic Engineering, Tsinghua University, China
JINGHUA PIAO, Department of Electronic Engineering, Tsinghua University, China
YUNKE ZHANG, Department of Electronic Engineering, Tsinghua University, China
YONG LI[*], Department of Electronic Engineering, Tsinghua University, China

In recent years, China has witnessed the proliferation and success of the online food delivery industry, an emerging type of the gig economy. Online food deliverers who deliver the food from restaurants to customers play a critical role in enabling this industry. Mediated by algorithms and coupled with interactions with multiple stakeholders, this emerging kind of labor has been taken by millions of people. In this paper, we present a mixed-methods analysis to investigate this labor of online food deliverers and uncover how the mediation of algorithms shapes it. Combining large-scale quantitative data-driven investigations of 100,000 deliverers' behavioral data with in-depth qualitative interviews with 15 online food deliverers, we demonstrate their working activities, identify how algorithms mediate their delivery procedures, and reveal how they perceive their relationships with different stakeholders as a result of their algorithm-mediated labor. Our findings provide important implications for enabling better experiences and more humanized labor of deliverers as well as workers in gig economies of similar kinds.

484

## 1 INTRODUCTION

The gig economy has been growing rapidly in recent years, where several representative platforms, e.g., Amazon Mechanical Turk [29], Uber [28], and Lyft [25], have achieved great success and have attracted the attention of a wide variety of researchers in the Computer-Supported Cooperative Work (CSCW) and Human-Computer Interaction (HCI) communities. Among these platforms are the ones for online food delivery services, which have been proliferating and burgeoning in recent

---

years in China. For example, on Meituan, one of the largest giants that center around these online food delivery services, more than 10 billion deals were completed in 2020, which contributes to a yearly transaction volume of more than 488 billion RMB[1].

A prominent factor that characterizes online food delivery services as part of the gig economy is their high reliance on the labor of online food deliverers: the services depend on the labor of these deliverers to dispatch the food prepared by merchants to the corresponding customers who order the food. This labor is vastly complex and intriguing through the lens of CSCW and social computing in two ways: 1) it involves frequent interactions with the mediation of algorithms, where online food deliverers receive both demands of dispatches and supports from the systems to complete their labor; and 2) it involves the frequent interactions with multiple stakeholders, where customers, merchants, platforms, deliverers, etc., all play indispensable roles in shaping this labor. Not only has this labor made possible the functioning of the services and thus the economy, but it has also provided job opportunities for millions of people and supported their living in China[2]. Given its growing essence in the Chinese society and the increasing societal impact it has been bringing upon, it is of great importance to understand how online food deliverers' labor is like in China. In light of this, in this paper, we conduct a nuanced investigation of how algorithms are mediating and shaping the labor of online food deliverers in China building upon prior preliminary steps towards this direction (e.g., [32, 40, 42]). Specifically, we explore the following research questions:

**RQ0**: How is labor like for online food deliverers and what factors are associated with their labor?

**RQ1**: How do algorithms mediate the labor of online food deliverers?

**RQ2**: How do online food deliverers perceive their relationships with related stakeholders and how are they shaped by the mediation of algorithms?

To answer them, we conduct a large-scale mixed-methods study to systematically understand people's practices and perceptions of labor as online food deliverers. Specifically, we take one of the duopoly online food delivery platforms as a case study because of their dominating position in the Chinese food delivery market (the duopoly of Meituan and Ele.me controls 90% of the market share [41]), burgeoning success, and representativeness of the typical platform design in the Chinese food delivery market. Quantitatively, we probe into the behavioral characteristics of deliverers with a data-driven approach. We leverage the records of 100,000 deliverers in Beijing to provide an objective overview of the working activities of online food deliverers. Qualitatively, we complement the quantitative analyses with an in-depth interview study with 15 deliverers in Beijing to further delve into deliverers' nuanced perceptions of how this labor is shaped by the mediation of algorithms. We identify in what ways and aspects algorithms are functioning throughout deliverers' labor, and analyze how deliverers perceive their relationships with related stakeholders as a result of their algorithm-mediated labor.

Our results show that 1) The labor of online food delivery services is highly flexible, time-dependent hour-wise, and can be shaped by many deliverer-related factors. 2) Algorithms mediate the labor of online food deliverers in work assignment, deadline setting, information support, and systemic ratings and judgments, which not only helps them better situate into their labor but also brings about brand-new challenges. 3) The algorithm-mediated labor of online food deliverers leads to their interesting relationships with related stakeholders, where those with customers are simple and ad hoc serving, those with merchants are symbiotic and mutually dependent, those with the platform are organized but not controlled, and those with other deliverers are like harmonious colleagues or even brothers and sisters. Based on our results, we discuss how to better serve and protect workers in online food delivery service platforms and similar platforms in the gig economy more broadly to foster more humanized labor and better experiences at work.

---

[1]http://media-meituan.todayir.com/202104190800003177739722495_tc.pdf
[2]https://mp.weixin.qq.com/s/3NwBL9zSX3nXWosTkSLhsQ

## 2 RELATED WORK AND BACKGROUND

In this section, we first introduce labor and the gig economy in CSCW, which our work situates as our context. We then review papers concerning algorithm mediation and algorithm management at work, which constitute our major aspects to delve into in this paper. We further discuss previous works on delivery work, highlight the research gaps and illustrate our contributions. We end this section by providing a background of online food deliverers in China, which we take as the case for study.

### 2.1 Labor, Gig Economy, and CSCW

Labor in work practices has long been concerned by Computer-Supported Cooperative Work (CSCW) and Human-Computer Interaction (HCI) researchers, where prior studies have been dedicated to uncovering the work practices of workers, identifying the impacts technology lays on labor issues, and proposing novel design practices to better support their labor [12]. Greenbaum [16] highlighted the necessity of including labor issues into CSCW. He discussed how incorporating labor issues such as wages, working conditions, and labor division could benefit CSCW and HCI design and illustrated how labor process analysis should be brought to help with design.

As a kind of platform that associates with "gig" by name, the gig economy has always been closely related to labor issues. Defined as "exchange of labour for money between individuals or companies via digital platforms that actively facilitate matching between providers and customers, on a short-term and payment by task basis" [6], the gig economy has received heated attention from the CSCW and HCI communities [8]. Based on how the gig work is completed, the gig economy can be broadly categorized into two categories: the virtual gig and the physical gig [5, 44, 47]. By referring to the notion of the virtual gig, it implies that the work can be completed online and anywhere; while as for the physical gig, it implies that the work needs to be accomplished offline and locally [5, 44, 47]. Representative examples of the virtual gig include microtask platforms such as Mechanical Turk [29] and paid playmate platforms [38], while typical scenarios of the physical gig include transportation services [12, 25, 28] (e.g., Uber, Lift, and DiDi in China), delivery services [36, 47] (e.g., Uber Eats, Deliveroo, and DoorDash and Meituan and Ele.me in China), and personal and household services [1, 44] (e.g., Taskrabbit and Gigwalk).

Numerous CSCW and HCI researchers have contributed their endeavors to understand labor in the gig economy. For example, in terms of the virtual gig, Martin et al. [29] analyzed how online crowdsourced turking serves as a form of invisible work, investigated Turkers' motivations, earnings, perceived relations with requesters, etc., and demonstrated practical and ethical issues associated with this invisible work. Hara et al. [19] extended them to analyze workers' earnings on Amazon Mechanical Turk, revealing the wage distribution and analyzing the causes of the low/high earnings. As for the physical gig, early studies focused on mobile crowdsourcing markets [20, 31], identified the labor dynamics and mobility patterns [31], analyzed how super agents stand out [31], and investigated the influence of worker situations such as busyness, fatigue, and presence of companions [20].

More recently, labor in the physical gig and on-demand mobile workforce of ridesharing has received much research attention [2, 12, 18, 24, 27, 28, 33]. For example, a first line of work focus on the delineation of the labor conditions. Glöss et al. [12] examined how ridesharing transforms the taxi business and brings brand-new labor opportunities and challenges through probing into multiple stakeholders' perspectives, where the emergence of emotional labor and worker flexibility are highlighted. Raval and Dourish [33] took a labor-practice-based approach to understand the emotional, body, and temporal labor in ridesharing and showed how it becomes laboring for ratings and calls for the transformation of space and the self. Ma et al. [28] extended them to leverage stakeholder theory to identify how to better serve and support Uber drivers and identified autonomy and earnings/expenses as two prominent stakes. Some works focused on the impact of algorithmic and data-driven management in shaping human labor in ridesharing services. For example, Lee et al. [25] investigated drivers' perceptions of the algorithmic work assignment, algorithmic information support, and algorithmic, data-driven evaluation systems. Based on the experiences and defects reflected by drivers, they provided design implications to better support ridesharing drivers. While some other studies focused on how the situated circumstances and contextual conditions shape drivers' labor, revealing how factors such as working part-time [27], regulations [2], and division of car owners and drivers [18] affect drivers' unique labor circumstances, experiences, and even social dynamics. However, contrary to the heated discussion on the virtual gig and the physical gig of transportation services and personal and household services by the CSCW and HCI communities, the physical gig of the delivery services has been understudied [36], where not until very recent years have preliminary steps been taken [36].

## 2.2 Algorithm Mediation and Algorithmic Management at Work

The rapid development of information technology has made possible the use of algorithms to support, mediate, and manage laborers at work [23]. Abundant research efforts have thus been laid on understanding how the mediation of the algorithms shapes organizational control, where Kellogg et al. [23] summarized the underlying mechanisms along the dimensions of "6Rs": the use of algorithms to *restrict* and *recommend* workers for direction, *record* and *rate* workers for evaluation, and *replace* and *reward* users for discipline. These algorithm mediation and management are especially prominent in the gig economy, which highly relies on algorithms to match tasks to crowdsourced laborers. For example, a frequently highlighted theme in prior literature is the control exerted through algorithm mediation and management in the gig economy [34]. As indicated by Rosenblat and Stark [35], these digital technologies and algorithms structure the information and power asymmetry between workers and the gig platform, resulting in the platform's control over laborers. However, this goes against workers' need for autonomy [30], which is exactly what workers call for in the scenario of gig work [28]. To resist this control, workers develop their own strategies to improve their power over their labor [30]. Other studies also identified how algorithmic management is achieved [25], what algorithmic management offers to workers [45], and even the benefits brought by the algorithmic control and surveillance in specific contexts [1].

## 2.3 Labor, Algorithm Mediation, and Delivery Work

Although relatively limited literature has focused on the gig of delivery work in the CSCW and HCI communities, there have been research attempts beyond these communities towards this direction. Specifically, most existing literature on labor in delivery work situates in Europe, Australia, and India [13, 17, 22, 36], where a lack of agency in schedule decisions has been expressed [22] and a sense of social isolation is reflected [36]. Economic security, work autonomy, and enjoyment are identified as important factors shaping deliverers' job quality [13], and safety-related concerns over the physical gig are highlighted [17]. As such, workers' mobilization for better wages and protections is on the rise [37, 43]. Similar but nuancedly different circumstances are also reflected in the Chinese context. For example, Sun [40] and Qiu et al. [32] identified how algorithms mediate aspects of temporality, emotional labor, and gamification that shape deliverers' everyday labor. Strong attraction, weak contract, intense regulation, and low resistance were identified as the main characteristics of deliverers' labor [48]. Deliverers demonstrate the trend of de-flexibilization, heading towards "sticky labor" that goes against the purported flexibility of the gig work as a result of income-related factors such as breadwinning [32, 42]. Deliverers also exercise "contingent agency" to work around their "structurally vulnerable position" in terms of platform capitalism [41]. Protests and disputes against their disadvantaged positions and seemingly unfair treatments can also be led to [41, 42].

However, despite the presence of the aforementioned lines of prior work, we identify research gaps that need to be filled. Firstly, most prior endeavors towards analyzing labor in delivery work rely only on self-reported data from surveys and interviews, which can suffer from selection bias and response bias. To tackle this, we extend prior work with a large-scale data-driven manifestation of the labor of online food deliverers (RQ0). Secondly, prior research into labor issues in delivery work in China mostly takes a critical stance and may even focus only on the negative aspects of this labor [3], which seldom offers a nuanced understanding of deliverers' situated feelings, thoughts, and reflections. We complement previous literature by illustrating how the mediation of algorithms brings both novel benefits and challenges to the labor of online food deliverers, where we seek to neutrally reflect deliverers' authentic labor (RQ1). Thirdly, although the influence of algorithm mediation on labor in the gig economy has been primarily examined, the related literature mostly focuses on the debate on whether gig workers should be identified as employees or technology consumers [14, 34], which seldom zooms in the intricate relationships between workers and *different* stakeholders as a result of the algorithm-mediated labor in detail. To address this, we seek to provide a detailed delineation of how deliverers perceive their relationships with related stakeholders (RQ2). Fourthly, the physical gig of online food delivery is relatively limited and understudied compared to its growing essence and prevalence. This circumstance is further exacerbated by the lack of CSCW and HCI research on the sharing economy and the gig economy in the Chinese context [8]. Therefore, we seek to further contribute to the CSCW and HCI literature with a nuanced understanding of the situated gig economy 1) in an understudied form and 2) in an underexplored culture by analyzing the algorithm-mediated labor of online food deliverers in China.

## 2.4 Case Study: Online Food Deliverers in China

The past few years have witnessed the proliferation and success of online food delivery services in China. For example, in 2020, more than 10 billion deals around online food delivery were made on Meituan, one of the duopoly online food delivery services that control 90% of the market share together with Ele.me [32, 41, 42], and the transaction volume reached more than 488 billion RMB, where the corresponding year-to-year growth achieved 16.3% and 24.5%, respectively[3]. We take one of the duopoly as a case study because of their dominating position in the Chinese food delivery market, their burgeoning success, and their representativeness of the typical platform design in the Chinese food delivery market.

Online food deliverers play a crucial role in enabling the success of online food delivery platforms: they are responsible for delivering the food from the merchants to customers and thus connecting the supply side and the demand side with their physical labor. Specifically, a wide variety of nearby merchants (restaurants, shops, etc. within about 3 kilometers) are aggregated on the platforms' apps. Customers can freely look through and place orders on these apps. A typical process of online food delivery starts when a customer finishes paying for the deal. The assignment algorithm allocates the deal to online food deliverers, where the information and location of the merchants and the customers are shown on another app especially for deliverers. Deliverers who are responsible for the deal head to the merchants to fetch the deals and bring them to the customers or the location assigned by customers, usually by an electric bike. After this, deliverers can click the button denoting service completion on their apps and a food delivery process is thus finished, where the deliverer will receive a certain amount of fee from the platform.

---

[3]http://media-meituan.todayir.com/202104190800003177739722495_tc.pdf

To better understand online food delivery services in China, we further highlight some prominent features concerning deliverers' work and thus their labor. Firstly, deliverers can be broadly categorized into two kinds: crowdsourced deliverers and specialized deliverers. Crowdsourced deliverers can freely choose whether to accept a deal assigned by the algorithm. They could go anywhere for online food delivery work and their delivery fee is based on how far their delivery distances are. Specialized deliverers are more constrained by the platforms. If they do not take a day off, they need to be online at the platform for at least a certain amount of time per day and have a minimum number of orders to deliver (which is often not hard to achieve). They are not able to reject the deals dispatched by the assignment algorithm. However, they have the right to shift the deals to other deliverers several times a day, where, if no one else chooses to accept the deals in several minutes, the delivery will go back to them again and they could not refuse. They only receive orders within a certain area (within several kilometers) and their delivery fee is given per deal. Secondly, different from other typical circumstances of the gig economy where only two-way relationships between customers and service providers are involved, three kinds of stakeholders, *i.e.*, merchants, deliverers, and customers, are present at the delivery procedure, which makes the process of online delivery more complicated and more nuanced. Thirdly, compared to traditional couriers whose delivery time is counted in days, the time allowed for online food delivery is a lot more limited to prevent the food from spoiling, which is often between 30 minutes and 60 minutes.

## 3 METHOD

### 3.1 Large-Scale Data-Driven Study

To gain an overall understanding of the labor of online food deliverers, we conduct a large-scale data-driven study to analyze their labor quantitatively. The dataset we use is collected from one of the duopoly that controls the market share of online food delivery services in China [32, 41, 42]. The dataset is derived from the all-round behavioral data of 100,000 randomly sampled deliverers who have made at least one delivery in Beijing from September 2021 to December 2021. Both specialized and crowdsourced deliverers were included in the dataset and were sampled based on their shares among all deliverers to better represent deliverers on the ground. The dataset is processed to contain 1) the overall distribution of deliveries and 2) highly-aggregated basic characteristics of these deliverers. For example, characteristics such as the average number of deals per day, the average lengths of delivery per day, the average distance they travel per day, on-time rate, etc. are recorded. Moreover, it is worth noting that similar to Yang et al. [46], we are not able to report the actual level of some dependent variables (especially the number of deals) due to confidentiality concerns for the benefits of the platform. Therefore, for those characteristics, we only report the normalized relative values rather than the absolute values.

*3.1.1 Ethical Considerations.* We take several careful steps to assure that privacy issues are well-protected concerning the sharing and mining of the data we use. Firstly, consent for research studies is included in the Terms of Service. Secondly, all identifiable personal information is removed and users' names and IDs are anonymized. Thirdly, all the collected data is stored in an offline server that is securely protected. Only authorized members of the research team are able to access the data, whose behaviors are also strictly bounded by non-disclosure agreements.

## 3.2 In-Depth Interview Study

We further complement the aforementioned large-scale data-driven study with an in-depth interview study, which was conducted between July and August 2021. To recruit participants, we first put up flyers at various places highly visible to online food deliverers. However, this proved to be ineffective because no online food deliverers contacted us through the contact information we left on the flyers. We thus changed our strategies for recruitment to directly ask free deliverers who were resting on the roadside and waiting for orders to be dispatched at several spots in person. In this way, we managed to recruit 15 online food deliverers of the same platform as in Section 3.1. We show the detailed information of our interview participants in Table 1. As we can see from the table, most of the interviewees were male, which accords with the fact that most online food deliverers are male [21]. We conducted the interviews in Mandarin in a face-to-face manner. Each of these interviews took 30-60 minutes and we compensated each of them for 80 RMB. All the interviews were audio-taped after we received oral consent from our participants. We transcribed them by combining the efforts of transcription services and manual modification, where we removed the identifiable information to protect our participants' privacy.

Table 1. Basic information of interviewees.

| Id | Gender | Age | Education | Type | Id | Gender | Age | Education | Type |
|---|---|---|---|---|---|---|---|---|---|
| P1 | M | 35-40 | Senior High School | Specialized | P2 | M | 20-25 | Technical High School | Specialized |
| P3 | M | 30-35 | Junior High School | Specialized | P4 | F | 25-30 | Technical High School | Specialized |
| P5 | M | 20-25 | Junior High School | Specialized | P6 | M | 25-30 | Senior High School | Specialized |
| P7 | M | 30-35 | Senior High School | Specialized | P8 | M | 30-35 | Junior High School | Specialized |
| P9 | F | 20-25 | Primary School | Specialized | P10 | M | 25-30 | Junior High School | Specialized |
| P11 | M | 25-30 | Bachelor's Degree | Crowdsourced | P12 | M | 30-35 | Junior High School | Crowdsourced |
| P13 | M | 30-35 | Bachelor's Degree | Crowdsourced | P14 | M | 30-35 | Junior High School | Crowdsourced |
| P15 | M | 25-30 | Technical High School | Crowdsourced | | | | | |

To analyze the data, we first adopted open coding [4] to the transcriptions. Three native Chinese authors independently analyzed and coded the first 20% of the transcriptions and gathered together to discuss the codes until complete consensus on the codes is reached. One of these authors then coded the rest of the transcriptions and constantly discussed with the other two authors whenever there were possibilities for uncertainty. Thereafter, the author translated the codes and the related quotes into English, where the other two authors were responsible for the verification of the translations. Upon finishing these steps, the whole research team met and thoroughly discussed the extracted content. We developed and kept refining the emerging themes through sub-categorization and constant comparison [39].

## 4 FINDINGS

### 4.1 Overall Working Activities

To characterize how labor is like for Chinese online food deliverers, we delve into our large-scale data-driven study to quantitatively delineate their working activities and identify what factors are associated with their labor.

In terms of the overall working activities, we first identify when the food delivery orders are completed. Figure 1(a) and Figure 1(b) show how the labor as online food deliverers vary across different days of the week and across different hours of the day on the platform. From the figures, we can see that deliverers' labor is time-variant. As shown in Figure 1(a), the number of deals completed grows from Monday to Saturday and drops a little bit on Sunday. Comparing weekends and weekdays, we find that more orders are completed on weekends than on weekdays: the number of deals completed on weekdays is 0.968 times the average, while for weekends the number rises to

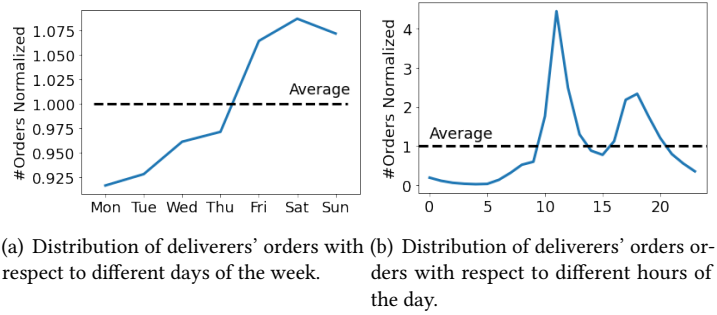

(a) Distribution of deliverers' orders with respect to different days of the week.

(b) Distribution of deliverers' orders orders with respect to different hours of the day.

Fig. 1. Relationships between time and deliverers' delivery.

1.080. However, this variance across different days of the week can seem small when compared with the variance across different hours of the day. As we show in Figure 1(b), deliverers' dispatches are highly time-depend at the hour level: the range of the average delivery deals across different hours is more than 4.4 times the mean of them. The average number of delivery deals rises gradually in the morning and sharply increases to its climax at 11-12 o'clock, where the peak volume can reach more than 4.4 times the average. When afternoon comes, deal number sharply reduces until about 15-16 o'clock, and then rises to its second peak around 18-19 o'clock, which is about 2.33 times the average. This pattern is easy to understand in that these two peaks denote people's lunchtime and dinnertime, which are exactly when customers are most likely to order food online. Moreover, compared to evenings when people can go home and cook dinner for themselves, when they are at work at noon, it would be less convenient for them to find food, which further makes online food delivery a good choice. As such, the labor of online food deliverers is highly time-depend hour-wise: in the morning and the afternoon, they can be relatively idle, but at the time for eating, especially for lunch, deliverers can be extremely busy.

We then analyze how the labor of online food delivery is completed by different kinds of deliverers in China. As manifested by the data, crowdsourced deliverers constitute 62.9% of all deliverers, while specialized deliverers take up a share of 37.1%. However, crowdsourced deliverers contribute to only 40.5% of all delivery orders on the platform, while specialized deliverers complete 59.5% of all delivery orders. The average number of orders that a specialized deliverer completes is 1.61 times the overall average, while the average number of orders that a crowdsourced deliverer completes is only 0.64 times the overall average. This demonstrates that the platform relies heavily on both crowdsourced deliverers and specialized deliverers to complete their services. Although there are more crowdsourced deliverers than specialized deliverers, the average number of orders that specialized deliverers complete is 2.5 times as large as that of crowdsourced deliverers. As a result, specialized deliverers are responsible for a larger share of the delivery of all orders.

Furthermore, we probe into the amount of work that Chinese online food deliverers complete per day per person, where we show the distribution of deliverers' daily working activities in Figure 2. For every deliverer, only the days that a deliverer has delivery records are regarded as the days on which they engage with the platform and are thus taken into the consideration of the calculation of averaging. For each figure, we use vertical dash lines to delineate the median values of the corresponding distributions. In terms of the number of orders deliverers complete per day, as shown in Figure 2(a), 90% of all deliverers carry out no more than 40 orders per day, and the median number of orders that deliverers take per day is 15-20. Specialized deliverers tend to shoulder more orders and share a median of 25-30 orders to deliver per day. While crowdsourced deliverers take

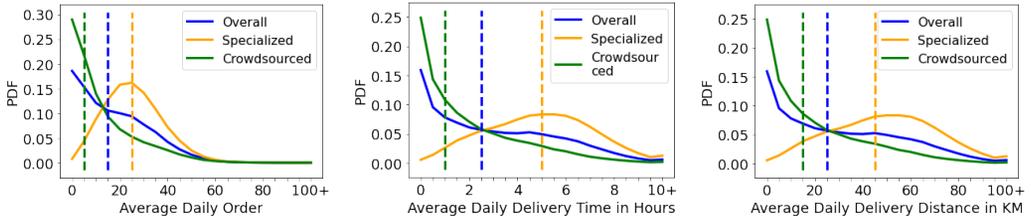

(a) Distribution of deliverers' average orders completed per day.

(b) Distribution of deliverers' average working hours per day.

(c) Distribution of the distances deliverers travel per day.

Fig. 2. Deliverers' daily working activities.

relatively fewer orders per day: almost 30% of them take only 0-5 orders per day, and the median number of deals that crowdsourced deliverers deliver is 5-10.

Another manifestation of online food deliverers' daily workload is the time they spend on the platforms per day. Here we investigate the overall distribution of online food deliverers' daily working hours, which is shown in Figure 2(b). From the figure, we can see that the vast majority of deliverers actually do not work so long a day on average: more than half of all deliverers work as an online food deliverer for 3 hours or fewer per day. However, if they would like to, they could choose to dispatch deliveries for even more than 10 hours per day. This flat and fat-tail distribution shows that deliverers have a large variance and thus flexibility over their labor in terms of how long they contribute to their labor. When penetrating into the work time of different kinds of deliverers, we find that specialized deliverers share a significantly higher median of delivery time, *i.e.*, 5-5.5 hours, while the median of crowdsourced deliverers is only 1-1.5 hours.

To study the efforts that online food deliverers put into their work, we probe into the average length that deliverers travel for their delivery daily. Our results indicate that deliverers truly have to travel much to complete the delivery work that they are responsible for (see Figure 2(c)). The median distance that deliverers travel per day is around 25-30 km. For specialized deliverers, this number can rise to as large as 45-50 km, while more than half of the crowdsourced deliverers travel less than 20 km per day. The flat and fat-tail distribution of deliverers' travel distance once again corroborates the large variance of the labor of online food delivery.

With these basic characteristics of deliverers' daily working activities identified, we further analyze how their labor associates with delivery performance related characteristics. We thus examine deliverers' labor with respect to their per-order delivery time and on-time rate (see Figure 3, where dash lines indicate the positions for the median). In terms of deliverers' average delivery time per deal, we can see from Figure 3(a) that most deliveries take less than 20 minutes, and deliverers share a median of 7.5-10 minutes for their average time for deliveries. Specialized deliverers' dispatches are more likely to take longer time (median: 10-12.5 minutes), while crowdsourced deliverers' average time for dispatches is relatively shorter (median: 7.5-10 minutes). We further investigate how deliverers' average number of daily orders for dispatches relates with their delivery time. From Figure 3(b), we can observe inverted U-shaped curves: when the average delivery time of a deliverer is extremely low, the average number of deals he/she delivers is not very high. It first rises when a user's average delivery time rises to 7.5-12.5 minutes, and then goes down significantly when deliverers' average per-deal delivery time rises to more than 20 minutes. The inverted U-shaped trend remains when only specialized deliverers or crowdsourced deliverers are considered, too.

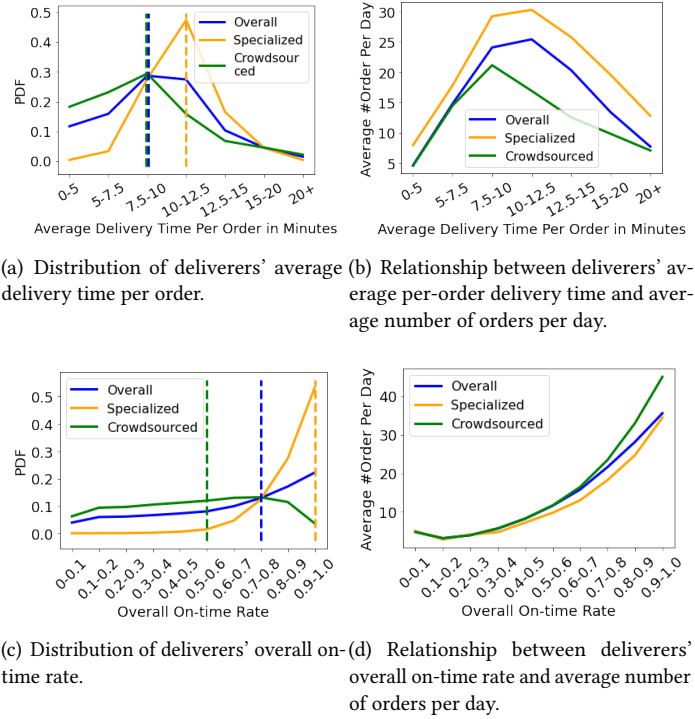

(a) Distribution of deliverers' average delivery time per order.

(b) Relationship between deliverers' average per-order delivery time and average number of orders per day.

(c) Distribution of deliverers' overall on-time rate.

(d) Relationship between deliverers' overall on-time rate and average number of orders per day.

Fig. 3. Deliverers' labor with respect to per-order delivery time and on-time rate.

As for deliverers' overall on-time rate, it can be seen from Figure 3(c) that online food deliverers' overall on-time rates are high (median: 0.7-0.8). Specialized deliverers' overall on-time rates are very high (median: 0.9-1.0), while crowdsourced deliverers' overall on-time rates are relatively lower (median: 0.5-0.6). Our further examination of how deliverers' number of daily orders associates with their overall on-time rate indicates a positive relationship between deliverers' number of daily orders and their overall on-time rate. As we delineate in Figure 3(d), deliverers' average number of daily orders increases from fewer than 5 to more than 35 when their overall on-time rate increases from 0 to 1. This trend holds for both specialized and crowdsourced deliverers (see Figure 3(d)).

In sum, as indicated by our data-driven investigations, Chinese online food deliverers' labor is highly time-dependent and flexible, where both the relatively more intensive labor of specialized deliverers and the relatively less intensive labor of crowdsourced deliverers constituent indispensable shares. This labor can be closely associated with deliverers' performance-related characteristics such as average per-order delivery time and overall on-time rate. However, despite this general understanding of the overall picture of online food deliverers' labor, it remains unclear how these deliverers encounter and perceive the nuanced labor-related circumstances. To provide more details on how this labor is like, we delve into results from our in-depth qualitative study, which we will show in the following sections.

### 4.2 Mediation of Algorithms

In this section, we investigate online food deliverers' experiences and reflections of their labor in detail based on our qualitative study. An emerging theme that shapes deliverers' labor throughout the whole process and lens that helps with a better understanding of deliverers' situated labor is the mediation of algorithms. Specifically, we find that with algorithmic work assignment, deadline setting, information support, and rating and judgment system, deliverers' labor is scaffolded and supported with new possibilities, yet brand-new challenges are brought upon along the way.

*4.2.1 Algorithmic work assignment.* One aspect of algorithm mediation that functions and shapes deliverers' labor at first sight is the work assignment algorithm. Whenever a customer places an order, it is up to this algorithm to dispatch the order to a nearby deliverer and to decide who is responsible for going to the merchants to fetch the meal and delivering it to the customer. As we have shown in our previous data-driven analysis in Section 4.1, the number of food delivery orders is highly time-dependent and centers around lunch and dinner time. This results in that at peak hours, the algorithm-supported work assignment system needs to dispatch several orders to a deliverer at a time to improve delivery efficiency and meet customer demand. As P5 says, *"In peak hours, we can get up to 8 or 9 orders at the same time. Under special circumstances, we will be assigned up to 12 orders"* (P5). In these circumstances, how these orders are dispatched would be among deliverers' top concerns.

For our interviewees, the concurrent work assignment system is identified as considerate in that it often assigns deliveries that are mostly locally concentrated or convenient for dispatches when they are assigned multiple orders. Specifically, as articulated by our interviewees, because the exact assignment logic remains a black box, they have developed "folk theories" [7, 10, 11] of the work assignment algorithm. According to their folk theories, closeness is a major mechanism they perceive to be governing the assignment algorithm. As expressed by P1, *"The system assigns orders based on closeness. If you are close to this merchant, it most likely assigns it to you."* (P1) This thus most likely leads them to be assigned a bundle of orders easy to reach, which vastly enhances their efficiency for delivery work. Even when deliveries are dispatched far away, the algorithmic assignment is perceived to have endeavored to be convenient for sequential dispatches. For example, P6 shares, *"If you are assigned an order at [place name] that is far away, when you are heading towards there, the system will mostly assign orders that are along your way there."* (P6) This has in turn made their long journey to the far-away places seem deserving and enhanced deliverers' perceived acceptance to these dispatches.

However, despite these, there can still be cases that the algorithmically assigned orders are non-ideal. Online food deliverers have developed their strategies in response to this. Specifically, one mechanism that almost all participants refer to is "order transfer". As designed by the platform, each deliverer can post three orders for help to a publicly visible space on the platform every day. If someone else responds within three minutes, the order is transferred to that person to be delivered. If nobody responds, the order will be returned to the initial deliverer. This mechanism allows orders to be distributed more rationally among deliverers. As explained by P10, *"If the order is farther away or I do not want to take it, I will check if it is on any people's delivery route. If so, they may take my order by the way"* (P10). In circumstances like this, whether there are idle deliverers and whether the dispatches would be convenient for other deliverers along the delivery of the orders at their hands can be decisive in shaping whether an order can be successfully transferred.

Some deliverers also figure out their tactics to avoid unwanted dispatches during peak hours. For example, P1 chooses to stay in a fixed area with numerous merchants, which is believed to have more orders. As P1 says, *"I just stay here habitually. There are more merchants and a higher likelihood of orders assigned in this area"* (P1). Some deliverers choose to turn off the order assignment system when unwanted orders are assigned (P2) or when too many orders are dispatched (P6). At this point, according to them, the algorithm will no longer assign orders to them and they only have to deliver previous orders. For some deliverers, this helps them avoid the dispatches of further undesirable orders. As mentioned by P2, *"During the peak hours, if you are assigned an order in a not very good location, you can get offline and hurry back after finishing this order. Then you can wait to be assigned with a better order"* (P2). Under this circumstance, deliverers feel that they can avoid being assigned more time-consuming orders and have easier orders to be assigned. For other deliverers, they turn off the assignment system to avoid order overload: *"If too many orders flood in, we have to knock off temporarily so that the system can no longer dispatch new orders ... With too many orders assigned, you are very likely to be late (for the deliveries)."* (P6)

*4.2.2 Algorithmic deadline setting.* Another aspect of algorithm mediation that shapes the labor of online food deliverers across the whole process of the delivery experiences is the deadline setting. Specifically, every time a customer places an order, the system sets a deadline for the deliverer to deliver each meal that is automatically calculated by algorithms, and the customer can see the estimated delivery time (which is close to this deadline) on their apps. For deliverers, they often try their best to keep to the deadline set by algorithms because 1) as we have shown in our previous data-driven investigation in Section 4.1, the on-time rate of a deliverer's deliveries can be an essential statistic that influences his or her labor in the number of orders dispatched to them, which is also what is believed by deliverers according to their folk theory (P1, P3, P6, P7) and 2) orders that overrun are also more likely to lead to negative reviews from customers, which could yield penalty for deliverers (P2, P7, P8). As P7 says, *"When you are overdue, customers are more likely to give you negative reviews. In this business, every deliverer does not want to have timeouts and negative reviews. They mean financial losses."* (P7)

Typically, the deadline for the delivery of each order set by the algorithm is about 30 minutes after the customer places the order and is automatically adjusted by algorithms. For example, the number of orders at hand is claimed to be a reason for the extension of deadlines by our interviewees (P7, P10). The more delivery tasks assigned to a deliverer at the same time, the longer the deadline of the newly dispatched delivery tasks will be: *"If there are seven or eight orders at the same time, the time limit would not all be half an hour. There are several orders that will extend the time limit, maybe up to an hour. If this were not the case, it would be impossible for us to deliver these orders on time. This is reasonable."* (P7) Furthermore, bad weather is also identified as a reason for the algorithm to extend the time limit by our interviewees: *"When the weather is bad, such as when it rains, there will be about half an hour more time than in normal weather for delivery. On rainy and snowy days, you cannot ride your electric bike too fast. Had the system not given me more time, I would not have been able to complete my delivery on time."* (P4) Similar circumstances are also mentioned by other participants such as P1: *"When it rains, we deliver more slowly. But the system gives us more time and the time it gives us is usually enough."* (P1) These case-dependent adjustments of deadline setting are appreciated by our interviewees, making them feel the system as humanized.

However, despite these highly-praised adjustments for time limits, there are cases where the deliverers do need more time to deliver, yet the system still keeps the typical time limits. The most commonly-mentioned situation is when merchants/restaurants are preparing food at too slow a speed, which is mentioned by all our interviewees. Specifically, the deadline given to the deliverer is timed from the moment the customer places the order. If the merchant/restaurant takes a long time to get the food prepared, relatively limited time will be left for deliverers' dispatches. As expressed by P9, *"Some merchants take too long to get the food prepared. I can't understand why some merchants are not even able to prepare the food within 30 minutes. Any time left for us for our delivery?"* (P9) To deal with this, the platform introduces a reporting mechanism that makes it possible to extend the time limits: *"If a restaurant is too slow in making food, the deliverer can report it to the platform up to twice, and each time the delivery time limit for this order can be extended by 6 minutes."* (P8) However, sometimes, especially in peak hours, the reporting mechanism fails to solve all the problems deliverers face: *"In busy periods, I am assigned several orders at a time, and I have to pick them all up before I can leave to deliver them. If a restaurant is too slow in preparing food for an order, I can report it to extend the delivery time for that order, but not for several other orders. It is unreasonable that a restaurant making food too slowly would prevent several of my orders from being delivered on time."* (P2) In cases like this, although the reporting mechanism allows the extension of time limits, the extension is only given to the specific reported order. However, at peak hours, deliverers can have several orders at hand. The time delay of a certain order can influence the delivery of other orders whose time limits remain the same, which thus results in failure to finish them on time.

Furthermore, deliverers may also be delayed in their journey to deliver food to customers because of physical constraints. For example, some neighborhoods, universities, or schools do not allow riding electric bikes. Deliverers thus have to complete their deliveries on foot, which can cause them to spend more time and lead to timeouts. P4 describes one past delivery experience: *"The neighborhood did not allow the entry of electric bikes, and I had a lot of things to deliver. I had to walk to the customer in two separate deliveries, which took a lot of time and ended up over time."* (P4) Sometimes elevators could also be sources of timeouts: to deliver food to customers in tall buildings, it often takes a long time to wait for the elevator. As P4 expresses, *"Some buildings have only one elevator available for deliverers. With so many deliverers at peak times, you have to wait in line for a long time to get upstairs and your order is definitely overdue."* (P4) To save time, some deliverers even choose to walk up and down the stairs in these circumstances: *"It would take a long time for me to get on the elevator, and if I did, the other orders I had on hand might overrun. So my only choice was to climb the stairs. There were times when I had to climb the stairs of more than 20 floors to deliver the food. This kind of order helps me lose weight, haha."* (P7) However, none of our interviewees mentioned that the time limits could be extended in these circumstances.

*4.2.3 Algorithmic information support.* Algorithms also step in to provide information support for online food deliverers to better complete their labor of deliveries. Considering the high dependence of deliverers' picking up food from restaurants and delivering it to customers on physical movements, knowledge of geographical characteristics, e.g., the location of restaurants and customers, is essential for deliverers to complete their delivery process. To fulfill these needs, the platform provides algorithmic support for spatial information. Specifically, the map system, the navigation system, and the route planning function are frequently mentioned by our interviewees, which are regarded as essential to help them complete their labor of online delivery services.

The built-in map system provides the location information of geographic entities, helping with locating the customers and the deliverers themselves. This help on localization is highly praised by deliverers: *"There are places that I don't know where they are. I can look at the map in the app. The map also shows me where I am now. It is very convenient."* (P4) This kind of information is especially important when the deliverers are not familiar enough with the locations, e.g., when they have just got to the city of Beijing and have just started their labor as online food deliverers. With the help of the map system, their unfamiliarity with the city and the locations is compensated, which in turn makes it possible for them to successfully complete their labor. However, sometimes the information provided by the map system can be insufficient and fails to cater to deliverers' situated needs. For example, sometimes the location information is limited to the building a place is located in rather than concretely providing where it is inside the building. This leads to circumstances that even though a deliverer knows which building it is in, it is not easy for him/her to find the merchants/restaurants: *"For example, the map only says that the restaurant is in [building name], but there are too many restaurants in [building name]. I have to ask another experienced deliverer to find it quickly."* (P6)

The navigation system is also frequently identified as a source of algorithmic information support for online food deliverers, which effectively provides route suggestions when a deliverer needs to get from one location to another in time. Considering the high reliance of online food deliverers' labor on movements and mobility, the navigation is of great help to deliverers, especially when they are novices: *"When you are new to the area, you need to rely on the navigation system to tell you how to get to the place. After all, you are not local."* (P3) Similar to the map system, the navigation system compensates for their lack of knowledge about the city and the locations when the deliverers are non-local, which is exactly the case for all our interviewees. Although it is not mandatory to take the recommendations, when deliverers have not accumulated much experience, they tend to follow the routes. The navigation system is also highly praised if the deliverers are poor in their sense of direction: *"I have a poor sense of direction, and navigation is necessary for me. Without navigation, I might take a long detour or have trouble finding my way and not be able to deliver on time."* (P10) For them, the navigation system can act as a crucial component guaranteeing that they could satisfactorily finish their labor. However, as mentioned by six of our interviewees (P2, P3, P6-P9), the essence of the navigation system can decay over time, where the system seems less necessary as the deliverers grow familiar with the roads and locations: *"You get familiar with it as time passes by and after that, you just need to look at the map to know the location, and you know how to get there."* (P3) For them, they have gradually developed their own knowledge of the regions, which lends them the confidence to get rid of the navigation system.

Algorithmic information support also comes from the route planning system. Sometimes, especially at peak hours, deliverers are responsible for delivering more than one order at the same time. In these circumstances, reasonable route planning can be extremely important, where built-in route planning stands out in helping delivers complete their delivery tasks more efficiently and avoid timeouts. As P4 explains, *"There are times when I am assigned many orders and it would be hard for me to decide which one to deliver first. The system will help me plan the route. By following this route, I usually don't exceed the time limit, which makes me feel easy."* (P4) Indeed, the deadline for each delivery can differ, and planning a proper delivery routine that does not exceed the time limit for every delivery can be tough work, especially when there are multiple deals at the hands of the deliverers at a time. The specifically designed route planning system takes the deadline for each delivery into account and provides an algorithmically optimal order for the deliveries, which can be heavily referred to and relied on. Therefore, although bearing the flexibility to determine in what sequence they complete the orders, deliverers such as P4 tend to follow the instruction given by the route planning system to avoid timeouts so as to better manage their deliveries and more satisfactorily complete their labor in time. However, further improvements on the route planning algorithms are also anticipated by our interviewees. For example, P8 argues that the algorithm for route planning fails to consider some situated obstacles: *"It only considers distance. It thinks it is convenient to cross the railway line, but in fact, it requires a large detour."* (P8) In circumstances like this, the algorithm only takes spatial distance into account and does not consider the actual time it takes to get from one location to another due to contextual barriers in their route planning. This could cast many difficulties in deliverers' dispatches and can result in unexpected timeouts, which are exactly what deliverers try their best to avoid.

*4.2.4 Algorithmic rating and judgment system.* Algorithms also mediate online food deliverers' labor through the rating and judgment system. Specifically, similar to prior sections, because factors such as how deliverers are actually evaluated and rated, what influence these ratings/evaluations would pose on deliverers remains a black box for deliverers, they once again rely on their own "folk theories" [7, 10, 11] to understand the algorithms and react to them. For example, four deliverers (P1, P3, P10, P11) mention that they believe indicators such as their average delivery time, on-time delivery rate, etc., will affect the number of orders dispatched by the assignment algorithm, which to a certain degree is corroborated by our previous quantitative analysis in Section 4.1: *"Your average delivery time per order is half an hour, while mine is 10 minutes. You are often late for the deliveries, while I never exceed the time limits. The system will certainly regard me as a more reliable deliverer and will give me more orders. There are times when others can only be idle and the system gives me orders. I can make more money."* (P3) This in turn encourages online food deliverers to behave in the way that they believe would result in being better treated by the algorithms, e.g., reducing their delivery time per order and accepting more orders, and try their best to improve their delivery skills.

Furthermore, according to our interviewees, the algorithm-mediated judgment system is responsible for figuring out the wrongdoers of customers' negative reviews. Specifically, when a negative review of an order is created by a customer, the system will decide why the negative review is left and who should account for it. If it was identified as irrelevant to deliverers' faults, deliverers would be freed from punishments. If it is due to a deliverers' fault, it can lead to a fine of 50 RMB to 100 RMB, which can be extremely disappointing for deliverers (P5-P10): as articulated by P8, *"A negative review will lead the platform to fine us 100 RMB, which means I will have to run a dozen of deliveries to earn it back"* (P8). Fortunately, the platform has left the space for deliverers to make appeals when they think they are mistakenly judged. P3 recalls one previous delivery experience: *"Once I was delivering food but the building the customer was in did not allow deliverers to go upstairs. The customer insisted that I go upstairs, but I couldn't. I put the food in the locker downstairs and told the customer to pick it up by himself. The system judged that it was my fault. But it wasn't. So I issued an appeal and the platform removed the review."* (P3) In these cases, when a deliverer issues an appeal, the system together with the corresponding staff will redetermine if the deliverer is to blame. If not, the penalties can be removed. According to some interviewees, the appeal mechanism can satisfactorily make fair judgments (P2, P3, P9, P10), which consequently guarantees their benefits to a certain degree. As articulated by P2, *"In most cases, the appeal system can make the right judgments. If it is indeed not the deliverer's fault, it can exempt you from the punishments."* (P2) However, this is not to say there are not circumstances when deliverers are still fined even when they think it is not their fault: *"I got a negative review even though it was clearly the restaurant's responsibility, and my appeal didn't work. I do not know why the system thinks it is my responsibility."* (P8) Cases like this can upset deliverers a lot and are extremely detrimental to deliverers' overall impressions of the rating system. Sometimes even when it is the deliverers' fault, the system seems to have too little tolerance for deliverers' accidental mistakes, especially considering the harshness of the punishments. For example, occasionally deliverers may claim delivery completion before actual delivery by accident, which can cause huge fines. According to P10, *"If we click 'delivery completion' by accident, it would cause much trouble, fining us several hundred RMB for punishment. We did not mean to break to rules, but there is no option for revocation."* (P10) In circumstances like this, possibilities to make up for the mistaken behaviors rather than directly giving huge fines are called upon.

## 4.3 Relationships with Related Stakeholders

To better understand the labor of online food deliverers, we further investigate how they perceive their relationships with different related stakeholders and how these relationships are shaped by and are results of algorithm mediation and the algorithm-mediated labor. As frequently mentioned by online food deliverers in our interview study, their labor is closely related to four kinds of stakeholders: customers, merchants, the platform, and other peer deliverers.

*4.3.1 Relationship with customers: simple and ad hoc serving.* Customers are one of the stakeholders that online food deliverers most frequently interact with: providing customers with quality food delivery service in time is identified as their main duty for their labor. As mentioned by P10, *"Our job is to deliver food to customers when they need takeout service."* (P10)

The mediation of the order-dispatching algorithm simplifies the relationship between a deliverer and a single customer, turning it to be ad hoc and temporal. As explained by P8, *"In a specific area, there could be dozens of or hundreds of deliverers, where there is the possibility that an order could be dispatched to all of us"* (P8). Therefore, compared with the case of logistics where a courier is responsible for an area and thus the courier could frequently interact with a customer, the interactions between an online food deliverer and a customer is more usually a one-time event: it starts when the algorithm distributes a customer's order to a deliverer and ends when the order is completed. They only need to interact with a specific customer once when handling his/her order, where they mostly only meet the customer once in one minute or so when handing the deliverers to them. This kind of algorithm-mediated short-term and ad hoc serving relationship alleviates their feelings of subordination. As expressed by P5, *"Different from being a waiter, being a deliverer makes me feel free from dealing with customers continuously and intensively"* (P5). This consequently makes some deliverers, e.g., P6, feel equal and reciprocal with their customers.

The algorithm-mediated simple and ad hoc serving relationship also shapes the style of the communication between deliverers and customers. Most deliverers (as mentioned by 11 out of our 15 interviewees) call their customers upon arrival or a few minutes before finishing the delivery to reduce the time spent in waiting for customers. For example, as expressed by P2, *"When I am about to arrive, I call customers and ask them to come down early to pick the food up or negotiate with them where I should put the food"* (P2). These communications are often brief and follow a concise pattern: deliverers first make a brief greeting, remind customers of the expected arrival time, and request customers to pick up the delivery at the appointed place and time. Moreover, most customers have a certain degree of consideration and respect for the deliverers. According to P13, *"When customers receive my calls, the majority of them are polite and keep their words"* (P13). Sometimes customers' words can even warm deliverers' hearts: *"They sometimes praise me and care for me when it is raining, when the package is heavy, or when I deliver it very fast"* (P9). This makes deliverers feel a sense of respect, which contributes to deliverers' praises of the relationships.

However, the relationship and communication patterns could sometimes be over-simplified and could also bring troubles to deliverers. P8 recalls one troublesome experience, *"When I arrived at the agreed location, no one picked up (the food). And the customer left a comment of 'no call, no knocking' (on the order recipe). So I took a picture for him (the customer), and I left, rushing for the next order ... But (the customer) claimed he didn't receive the food ... I felt helpless and aggrieved"* (P8). Similar cases are also mentioned by three other participants (P3, P5, P7). Although these troublesome cases do not occur frequently, they are detrimental to deliverers' overall working experiences. Faced with these troubles, deliverers tend to first explain their situations to customers, and then request for understanding, and even compensate for the missing food or broken packages (P4, P5, P8, P9). As shared by P9, *"When the food is lost, I will call the customers to explain the situation and hope they will forgive me. But most of the time I will pay for the lost food to reduce the trouble"* (P9). Despite these efforts, sometimes customers still leave negative reviews on the platform, which might lead to very high fines and affect the number of orders assigned by algorithms according to our interviewees: *"One negative review will lead to a penalty which is comparable to the income from dozens of orders"* (P8); *"(negative reviews) will affect the order-dispatching algorithm"* (P1, P10).

*4.3.2 Relationship with merchants: symbiotic and mutually dependent.* Merchants are another kind of stakeholder that online food deliverers could not avoid interacting with: only after fetching food from the merchants can deliverers start their delivery. Regarding their relationship with merchants, online food deliverers regard it as *"symbiotic, cooperative, and mutually beneficial"* (P9). As explained by P9, *"without my delivery, their food could not be handed to the customer; without their presence, I would have nothing to deliver"* (P9). As such, online food deliverers and merchants are symbiotic, mutually dependent upon each other and beneficial to each other. When these symbiotic and mutually beneficial interactions become frequent (because as we have mentioned, some specialized deliverers are only responsible for certain areas), their relationship can be further enhanced. For example, P2 mentions, *"We get very familiar and would occasionally chat with each other when we collaborate frequently"* (P2).

The mutual dependency between merchants and deliverers is further strengthened by the algorithmic deadline setting of the system. According to the deadline setting algorithm, the time given for merchants' preparations for food and deliverers' dispatches are counted as a whole: *"we [the merchant and the deliverer] share an amount of time, for example, 30 minutes, for cooking and delivery ... If it takes 5 minutes for him/her to cook, then I will have 25 minutes for my delivery. But if he/she uses 29 minutes, then I will only have 1 minute, which will definitely make me run out of time"* (P9). As such, the logic of the algorithm makes an on-time delivery a close collaboration between merchants and deliverers, where the performance of one side can directly affect the other. For example, from the deliverers' perspective, as shown in the aforementioned example, the merchant's preparation time directly determines when an order can be delivered and even affects the timing of the following orders. The consequence of merchant's failure to get the food prepared in time can be detrimental to deliverers. As reflected by P14, *"If a merchant cannot finish their preparation for the food, the deliverer has to wait for it and wastes lots of time. And the following orders will be delayed, too."* (P14) Fortunately, these cases do not happen so often, where most merchants leave enough time for deliverers to finish the delivery in time (P1, P3). As explained by P1, *"Merchants also know that we should not be late for deliveries and will give us the food as soon as possible. In some cases, merchants will even let deliverers who are in a hurry cut in line"* (P1). Even when cases when merchants prepare the meals too slowly happen, the system allows deliverers to get rid of punishments by introducing a reporting mechanism: *"If a merchant prepares a meal too slowly, you can report it every 6 minutes provided that you are at the merchant's place. After you have reported it twice, timeouts would not be identified as your fault."* (P3)

*4.3.3 Relationship with the platform: organized but not controlled.* In terms of the platforms, ten of our interviewee deliverers mention that they are well-organized by the platform and its grassroots organizations – stations or *zhandian* in Chinese (P1-P10). As expressed by P4, *"All we specialized deliverers belong to stations, and we are organized by the platform through these stations."* (P4) Specifically, deliverers are partitioned to different local stations based on their respective delivery areas (P10). For common deliverers, the local station is the main channel for the organization of the platform-related affairs. For example, deliverers recall they are trained by their stations when they are newcomers. As introduced by P4, *"Before we start our work, (the station) will clarify what we should do, for example, (the station) tells us to wear work uniforms, masks, and helmets, and what we shouldn't do, ... for example, not to cheat on the delivery."* (P4) Some stations hold morning meetings with their delivery staff, emphasizing the importance of safety and sharing delivery experiences. As reflected by P7, *"On the morning meeting, (the station) tells us about the issues that happened the day before, and summarize experience ... (The station) emphasizes the importance of safety, including the safety of us and the delivered food."* (P7)

The organization of the stations is often loose, which lends deliverers a high degree of freedom and autonomy according to our interviewees (P2, P7, P9). Rather than superiors to strictly conform to and to be controlled by, the stations and stationary staff are more often regarded as resources for help. As mentioned by P2, *"The existence of the stations is good ... Except for training and meetings, the station does not control us. Instead, I think it is mostly helping us ... For example, when I am met with an accident, I will turn to the station for help."* (P2) This is especially the case when deliverers encounter hardships with the algorithm mediation. For example, as P3 and P7 express, *"The stations helps us by re-distributing orders to other deliverers when we are assigned too many deliveries to dispatch"* (P3). *"They help us mediate problems, for example, negative reviews, among users, the platform, and me"* (P7). Moreover, as mentioned by P9, *"They sometimes give us bonuses"* (P9). The harmonious relationships between these stations and deliverers make some deliverers even regard staff at the stations as their bros (P2, P7, P9). As explained by P7, *"(the station head and I) are all workers ... We work, eat, chat, and deal with troubles together. They are more like my friends or bros, rather than my leaders."* (P7)

*4.3.4 Relationship with other deliverers: harmonious colleagues or even brothers and sisters.* Although there is seldom direct interaction among online food deliverers concerning their labor during a single delivery, their shared identity as online food deliverers can bring them together, especially for the ones whose areas of delivery are similar. As reflected by P7, *"We are all deliverers ... We deliverers within the same region or sometimes even across different regions will greet each other when we meet. We act as if we were acquaintants."* (P7) This friendly atmosphere, as further explained by P7, is partly a result of the assignment logic of the algorithms: *"We all do our own work assigned by the algorithm and there is no competition among us."* (P7) Indeed, the system assigns orders to deliverers rather than requiring deliverers to compete for orders. This assignment algorithm mediates deliverers' aggressive competition for orders, which set a harmonious tone for their relationships with each other. This harmonious atmosphere is further strengthened by the time-dependent nature of the labor of online food delivery, which leaves deliverers time for offline social interactions and conversations between busy periods. For example, as mentioned by P9, *"While waiting for order assignments, many of us share our lives. We talk about various topics, like games, food, and places to stay"* (P9). Some experienced deliverers also share their tricks, which gives rise to further mutual benefits. As P2 says, *"We share with each other where the hidden shortcuts are, where the entrance to a particular neighborhood is, and which merchant is harsh"* (P2).

With these friendly interactions instead of fierce competition, some deliverers even become brothers and sisters as mentioned by four of our interviewees (P2, P3, P5, P9). For example, P5 recalls that *"Sometimes I am too busy to pick up takeouts from distant merchants, so I will ask my bros to help me if any of them passes by the merchant"* (P5). With these communities of brothers and sisters who communicate through WeChat groups and direct messages, community members can easily get support for their work. These communities can be extremely important when the deliverers are women, where their help could support women deliverers in more smoothly engaging in the work of delivery. As expressed by P4, *"Once the delivered item was too heavy to carry for me, (one of my peer deliverers) came to my help, thus preventing me from having to quit the job"* (P4). Moreover, when deliverers are increasingly familiar with each other, they even occasionally get together for dinner to further consolidate their relationships (P2, P3): *"For those (peer workers) we get along well with through our work, we sometimes go out for a drink on weekends."* (P3)

## 5 DISCUSSIONS

As shown in our results, the labor of online food deliverers in China demonstrates distinct characteristics, bearing a certain degree of similarities as well as differences with its prior counterparts in the gig economy. In this section, we first tease out these similarities and differences, and then go on to discuss how the mediation of and interactions with algorithms shape deliverers' working experiences. We further illustrate the implications that could be drawn from the current work. We end this section by discussing the limitations and future work of our study.

### 5.1 Online Food Delivery Services and Other Gig Economy

As an emerging kind of platform that fits into the gig economy, online food delivery services share many similarities with other gig economies. This is manifested throughout our work, where we take one of the dominating online food delivery platforms in China as a case study. For example, aligning with prior literature which indicates the merits of worker flexibility for on-demand work in general [14] and ridesharing platforms such as Uber [12, 34] specifically, we discover that flexibility is also enjoyed by deliverers of online delivery services in China. They share a degree of freedom in deciding how long they work for a day at their will, where their labor can be highly influenced by performance-related factors such as average per-order delivery time and the overall on-time rate. This close link between labor and these performance-related factors once again corroborates what has been documented by prior works on other forms of the gig economy [14, 25, 33, 34].

However, online food delivery services also have their own uniqueness. Firstly, as we have manifested, the orders on these platforms are highly time-dependent hourly compared with express couriers. Considering the time-dependent nature of people's eating habits and the fact that deliverers need to send the food to customers in a short amount of time, the labor of these deliverers is thus also highly time-dependent hourly, where a significant portion of labor centers around peak hours at noon and in the evening. This means that the demand for the labor of online delivery services is not equally distributed temporally, which leaves sufficient possibilities for part-time labor. Secondly, online food delivery services are especially distinctive in that usually three kinds of stakeholders are engaged in a typical dispatch. Compared to crowdsourcing platforms where on most occasions only requester-worker relationships are present [29] and ridesharing platforms where mostly only driver-customer relationships are considered [12, 33], the circumstances of online food deliverers are more intricate: the performance of the newly incorporated stakeholder, merchants, can directly influence deliverers' labor, where their mutual understanding and collaborations are highlighted. This underscores the importance of tackling the unique challenges posed by the more complicated relationships. Thirdly, the interactions between online food deliverers and customers are rather ad hoc and are often only restricted to a single encounter. Compared to couriers when a deliverer is responsible for an area, dozens or even hundreds of online food deliverers all have the possibility of being assigned a deal, which makes it hard for the establishment of long-lasting relationships. Therefore, although emotional labor is still present for deliverers to earn customers' positive evaluations [32, 40] considering the essence of the customer-facing algorithmic rating system in shaping deliverers' labor, the relatively limited amount of time in the interactions between online food deliverers and customers results in relatively less demand on emotional labor in food delivery services compared to other circumstances such as ridesharing [26, 33]. This further lowers down the requirements for the labor of online food deliverers and makes the labor seem easier to be adopted. Moreover, the relatively limited amount of interaction time between deliverers and customers also eliminates the possibilities of harassment and assaults reported in other physical gig platforms such as ridesharing [34]. This further helps avoid security-threatening worker-customer relationships and improve workers' overall working experiences.

### 5.2 Mediation of and Interactions with Algorithms in Human Labor

As we have demonstrated in our findings, a major aspect that shapes the labor of online food deliverers is the mediation of algorithms, where through deliverers' interactions with these algorithms, new benefits and challenges emerge. Specifically, most prior literature focused on how platforms exert control over laborers through algorithmic managements [14, 23, 34, 35]. However, as we have shown in our qualitative study, this notion of "control" may not reflect deliverers' situated reflections towards how algorithms are actually mediating their labor. Indeed, algorithms "support" deliverers' labor in various ways, which are appreciated by the deliverers; they also reduce hierarchical human control, which lends deliverers a higher degree of flexibility and freedom perceived. However, this is not to say that the mediation of algorithms in the gig work of online food delivery is good in all aspects. It gives rise to brand-new challenges along the way, e.g., failure to take certain physical constraints into account, which remains for future design to address so as to better protect deliverers' stakes.

To authentically reflect human labor under the mediation of algorithms in the scenario of online food delivery in China, we conduct a mixed-methods analysis combining a large-scale quantitative study with in-depth qualitative interviews. To better outline the impact of algorithmic management on human workers, we take an objective stance, which is in line with Lee et al. [25] and some similar works. However, the concrete scenario of online food delivery can cultivate novel experiences that are beyond just being similar to these prior studies, especially when compared with ridesharing. For example, although information support is regarded as vital for labor in both ridesharing and online food delivery, in the scenario of ridesharing, information about surge prices is more valued [25], while in online food delivery services, emphasis is laid more on functional geographical information, e.g., locations. Moreover, the circumstance of online food delivery has also enabled new interaction experiences with new algorithms. A salient characteristic is the algorithmic deadline setting. Because of the reliance of online food delivery services on timeliness, issues related to time limits receive the attention of many or even most deliverers. When algorithmically determined deadline setting systems are introduced, brand-new advantages and concerns are brought upon, e.g., how to better manage the order of the deliveries to avoid timeouts and how to deal with unexpected timeouts.

Beyond the labor itself, the mediation of algorithms also vastly shapes online food deliverers' relationships with related stakeholders, which could be vital for a better understanding of the overall picture of labor in the online food delivery industry and thus the gig economy. Indeed, what the relationships are claimed to be and are supposed to be between on-demand gig platforms and workers has been heatedly discussed by scholars and attracted much public attention [14, 34]. Extending prior works that take a regulation- and accountability-based view to focus on distinguishing whether gig workers should be legally classified as employees or technology consumers [14, 34], we seek to delineate how Chinese online food deliverers authentically think of their relationships with different related stakeholders. As we have shown in Section 4.3, the algorithmic work assignment system simplifies the interactions between customers and deliverers, turning the relationships to be short-term and ad hoc. The algorithmic deadline setting system that combines merchants' preparing time and deliverers' delivery time together further makes the relationships between merchants and deliverers symbiotic and mutually dependent, where a higher degree of collaboration between these stakeholders is called upon to get the labor satisfactorily completed. The existence of the loosely-organized stations harmonizes the relationships between deliverers and the platform by acting as resources for help and managing deliverers' hardships. The algorithmic logic of the work assignment system that relies on systemic dispatches rather than competition casts a layer of harmony on deliverers' relationships with their peer workers, making it possible for the labor to be cooperative rather than competitive. In showing these, we show the possibilities of scaffolding and retaining newcomers, which could be a long-standing concern of the gig economy [14], with the combination of algorithmic support and human help from the grassroots stations. We shed light on the feasibility of relieving the tensions between gig workers and platforms (e.g., lack of effective approaches for appeals, which could be pervasive and detrimental [14]), by introducing self-organized managers or human coordinators. We accord with Gray et al.'s [14, 15] highlight of the essence of collaboration in the gig economy, while differ from Yao et al's [47] revelation of people's reluctance to share due to the fear of competitive disadvantages and limited empathy, which might be attributed to differences in the design of platform logic. Experiences of social isolation [36] are not reflected in our scenario, too, which may be the result of the presence of peers in the same stations, possibilities of offline social interactions and conversations between busy periods, or cultural differences. As such, we extend prior literature on the impact of algorithmic aspects on labor in the gig economy (e.g., Lee et al. [25]) to their impacts on laborers' relationships with different relevant stakeholders, which constitutes laborers' contextual working conditions and also plays important roles in shaping their overall experiences.

### 5.3 Implications

Our study also provides novel implications for related areas to further protect the interests and benefits of online food deliverers as well as workers in similar scenarios of the gig economy.

**Taking into account context-specific particularities.** One defect of the current algorithmic design is the failure to take into account context-specific particularities. For example, deliverers complain that the time limit set by the algorithms for delivery focuses too much on spacial distances and fails to consider the cases when 1) they need to take busy elevators for their delivery, and 2) they need to walk into some neighborhoods, universities or schools to complete the deals, which could consume much time and lead to timeouts. To design more human-centric algorithmically supported systems, practitioners should keep an eye on these practical barriers triggered by real-world particularities and lend workers a stronger sense of care.

**Making algorithmic mediation more transparent and understandable to the laypeople.** As we have shown in previous sections, the mediation of algorithms has brought upon not only opportunities but also concerns to online food deliverers. Some of these concerns can be attributed to the fact that the algorithms seem like "black boxes" for the deliverers: they could only rely on their own folk theories to understand the algorithms and self-hypothesized concerns could arise. To deal with this, we deem it vital to make the algorithms more transparent. Although some research endeavors have been dedicated towards this direction, they can be too hard for laypeople like deliverers to understand [9]. Therefore, we highlight the essence of providing explanations and enhancing the transparency of the algorithms in a way that deliverers could easily grasp. This could include not only delineating how they achieve them technologically, but also illustrating some of the mechanisms in words that they can comprehend.

**Improving deliverers' control over the systems.** Some of the current challenges articulated by deliverers can be ascribed to their relatively restricted control over the systems. For example, when a merchant/restaurant is too slow in preparing the food, deliverers could only report it in the system for the extension of time limits, but could hardly do anything to the misbehaved merchants. To better protect deliverers' benefits, practitioners could enable the deliverers to rate the merchants, where poorly rated merchants would be automatically assigned longer time limits by the system. As such, the deliverers can more easily deal with the cases, and for merchants, this can serve as a kind of punishment. Similarly, the fact that they could receive a huge monetary loss when mistakenly clicking the button of delivery completion could be ascribed to their lack of options for revocations. To deal with this, practitioners should consider incorporating these features to better protect the stakes of laborers.

### 5.4 Limitations and Future Work

There are also some limitations of our study, which leaves potential avenues for future work. Firstly, in terms of our interview study, there is the possibility of selection bias. Although we try to mitigate this by randomly finding participants at several sites, it is possible that the ones who agree to participate in our studies are generally more positive towards the platform. Secondly, our work is based on the case of one dominating online food delivery platform in the context of China. There is the possibility that the platform-specific design and the socio-cultural background of China could be sources of biases. However, its dominant position in constituting the duopoly of the online food delivery industry in China [32, 41, 42] ensures our study to enjoy a degree of representativeness of deliverers' labor in China, and the fact that the platform design across different platforms of the Chinese online food delivery services is similar [32, 40] further enhances the generalizability of our results in the Chinese context. Its current huge volume and the burgeoning of online food delivery service platforms in China also lead us to believe that the case itself is worth studying and exploring in depth. To further verify the robustness and generalizability of our results, we call on future studies to conduct cross-platform and cross-culture analyses, especially across platforms from different nations considering the similarity of platform design in the Chinese online food delivery industry. Thirdly, our work only examines the circumstances of the platform within a specific period of time. This could potentially induce biases as a result of the platform's specific stage of development. Therefore, future research can consider following Sun et al. [42] to attend to how circumstances concerning deliverers' labor change over time.

## 6 CONCLUSION

In this paper, we seek to understand the labor in a burgeoning branch of the gig economy – online food delivery platforms in China. Combining large-scale data-driven investigations and in-depth interviews, we seek to uncover the labor of online food deliverers whose endeavors are highly mediated by algorithms. Our results show the labor of online food deliverers is highly time-dependent, can be influenced by deliverer-related characteristics, and is vastly coupled with and mediated by algorithms, which helps deliverers better accommodate their work but also catalyzes new contextual obstacles. The algorithm-mediated labor also brings novel interaction experiences and cultivates deliverers' nuanced relationships with different stakeholders. Based on our discoveries, we further discuss how our work provides novel implications for enabling more human-centric labor of deliverers and even benefit workers in similar gig economy based platforms.

## ACKNOWLEDGMENTS

The authors thank Haoran Guo for his help on some of the interviews and Jing Yi Wang for her helpful suggestions on polishing the work. This work was supported in part by the National Key Research and Development Program of China under grant 2020YFA0711403 and the National Natural Science Foundation of China under U1936217, 61971267, and 61972223.

A Mixed-Methods Analysis of the Algorithm-Mediated Labor of Online Food Deliverers in China 484:23[6] United Kingdom Department for Business, Energy & Industrial Strategy. 2018. The Characteristics of Those in the Gig Economy. https://assets.publishing.service.gov.uk/government/uploads/system/uploads/attachment_data/file/687553/The_characteristics_of_those_in_the_gig_economy.pdf

[7] Michael A DeVito, Darren Gergle, and Jeremy Birnholtz. 2017. "Algorithms ruin everything" #RIPTwitter, Folk Theories, and Resistance to Algorithmic Change in Social Media. In *Proceedings of the 2017 CHI conference on human factors in computing systems*. Association for Computing Machinery, New York, NY, USA, 3163–3174.

[8] Tawanna R Dillahunt, Xinyi Wang, Earnest Wheeler, Hao Fei Cheng, Brent Hecht, and Haiyi Zhu. 2017. The sharing economy in computing: A systematic literature review. *Proceedings of the ACM on Human-Computer Interaction* 1, CSCW (2017), 1–26.

[9] Upol Ehsan, Samir Passi, Q Vera Liao, Larry Chan, I Lee, Michael Muller, and Mark O Riedl. 2021. The Who in Explainable AI: How AI Background Shapes Perceptions of AI Explanations. *arXiv preprint arXiv:2107.13509* (2021).

[10] Motahhare Eslami, Karrie Karahalios, Christian Sandvig, Kristen Vaccaro, Aimee Rickman, Kevin Hamilton, and Alex Kirlik. 2016. First I" like" it, then I hide it: Folk Theories of Social Feeds. In *Proceedings of the 2016 CHI conference on human factors in computing systems*. Association for Computing Machinery, New York, NY, USA, 2371–2382.

[11] Motahhare Eslami, Kristen Vaccaro, Min Kyung Lee, Amit Elazari Bar On, Eric Gilbert, and Karrie Karahalios. 2019. User attitudes towards algorithmic opacity and transparency in online reviewing platforms. In *Proceedings of the 2019 CHI Conference on Human Factors in Computing Systems*. Association for Computing Machinery, New York, NY, USA, 1–14.

[12] Mareike Glöss, Moira McGregor, and Barry Brown. 2016. Designing for labour: uber and the on-demand mobile workforce. In *Proceedings of the 2016 CHI conference on human factors in computing systems*. Association for Computing Machinery, New York, NY, USA, 1632–1643.

[13] Caleb Goods, Alex Veen, and Tom Barratt. 2019. "Is your gig any good?" Analysing job quality in the Australian platform-based food-delivery sector. *Journal of Industrial Relations* 61, 4 (2019), 502–527.

[14] Mary L Gray and Siddharth Suri. 2019. *Ghost work: How to stop Silicon Valley from building a new global underclass*. Eamon Dolan Books.

[15] Mary L Gray, Siddharth Suri, Syed Shoaib Ali, and Deepti Kulkarni. 2016. The crowd is a collaborative network. In *Proceedings of the 19th ACM conference on computer-supported cooperative work & social computing*. Association for Computing Machinery, New York, NY, USA, 134–147.

[16] Joan Greenbaum. 1996. Back to Labor: Returning to labor process discussions in the study of work. In *Proceedings of the 1996 ACM conference on Computer supported cooperative work*. Association for Computing Machinery, New York, NY, USA, 229–237.

[17] Denis Grégoire. 2017. Delivering for FoodTech: at your own risk. *HesaMag,(16)* (2017), 17–21.

[18] SM Taiabul Haque, Rayhan Rashed, Mehrab Bin Morshed, Md Main Uddin Rony, Naeemul Hassan, and Syed Ishtiaque Ahmed. 2021. Exploring the Tensions between the Owners and the Drivers of Uber Cars in Urban Bangladesh. *Proceedings of the ACM on Human-Computer Interaction* 5, CSCW1 (2021), 1–25.

[19] Kotaro Hara, Abigail Adams, Kristy Milland, Saiph Savage, Chris Callison-Burch, and Jeffrey P Bigham. 2018. A data-driven analysis of workers' earnings on Amazon Mechanical Turk. In *Proceedings of the 2018 CHI conference on human factors in computing systems*. Association for Computing Machinery, New York, NY, USA, 1–14.

[20] Kazushi Ikeda and Keiichiro Hoashi. 2017. Crowdsourcing go: Effect of worker situation on mobile crowdsourcing performance. In *Proceedings of the 2017 CHI Conference on Human Factors in Computing Systems*. Association for Computing Machinery, New York, NY, USA, 1142–1153.

[21] Meituan Research Institute. 2020. Report on Employment of Meituan Deliverers During 2019 and 2020 COVID-19 Outbreak. https://mp.weixin.qq.com/s/SLsOV4RUpKDmFnW8ZJOmVQ

[22] Mirela Ivanova, Joanna Bronowicka, Eva Kocher, and Anne Degner. 2018. *Foodora and Deliveroo: The App as a Boss? Control and autonomy in app-based management-the case of food delivery riders*. Technical Report. Working Paper Forschungsförderung.

[23] Katherine C Kellogg, Melissa A Valentine, and Angele Christin. 2020. Algorithms at work: The new contested terrain of control. *Academy of Management Annals* 14, 1 (2020), 366–410.

[24] Donghun Lee, Woochang Hyun, Jeongwoo Ryu, Woo Jung Lee, Wonjong Rhee, and Bongwon Suh. 2015. An analysis of social features associated with room sales of Airbnb. In *Proceedings of the 18th ACM conference companion on computer supported cooperative work & social computing*. Association for Computing Machinery, New York, NY, USA, 219–222.

[25] Min Kyung Lee, Daniel Kusbit, Evan Metsky, and Laura Dabbish. 2015. Working with machines: The impact of algorithmic and data-driven management on human workers. In *Proceedings of the 33rd annual ACM conference on human factors in computing systems*. Association for Computing Machinery, New York, NY, USA, 1603–1612.

[26] Christoph Lutz, Gemma Newlands, and Christian Fieseler. 2018. Emotional labor in the sharing economy. In *Proceedings of the 51st Hawaii international conference on system sciences*.Proc. ACM Hum.-Comput. Interact., Vol. 6, No. CSCW2, Article 484. Publication date: November 2022.